\documentclass[sigconf]{acmart}
\usepackage{subcaption}
\usepackage{multirow}
\usepackage[most]{tcolorbox}
\usepackage{booktabs}
\usepackage{etoolbox}
\newtcolorbox{promptbox}[1][title]{
    colback=gray!5,
    colframe=black!70,
    boxrule=0.6pt,
    arc=2pt,
    left=6pt,
    right=6pt,
    top=6pt,
    bottom=6pt,
    breakable,
    title={#1},
    fonttitle=\bfseries,
    before upper={\parindent0pt\obeylines}
}

\AtBeginDocument{
  }

\setcopyright{rightsretained}
\acmDOI{}
\acmISBN{}
\acmYear{2026}
\copyrightyear{}
\acmConference[EARL at CIKM '26]
  {3rd Workshop on Evaluating and Applying Recommender Systems
   with Large Language Models}
  {November 7--11, 2026}
  {Rome, Italy}

\makeatletter
\patchcmd{\maketitle}
  {\acmConference@shortname, \acmConference@venue}
  {\acmConference@name\space(\acmConference@shortname),
   \acmConference@date, \acmConference@venue}
  {}
  {\PackageError{acm-footer}{Could not patch conference footer}{}}
\makeatother

\begin{document}

\title{From a Large-Scale Teacher to Efficient Multimodal Students for Search Relevance Prediction}

\author{Tomasz Gniazdowski}
\authornote{Both authors contributed equally to this research.}
\email{tgniazdowski@ebay.com}
\affiliation{
  \institution{eBay}
  \city{}
  \country{}
}
\affiliation{
  \institution{Tooploox}
  \city{Wroclaw}
  \country{Poland}
}
\author{Reza Momenifar}
\authornotemark[1]
\email{rmomenifar@ebay.com}
\affiliation{
  \institution{eBay}
  \city{San Jose}
  \country{USA}
}
\author{Nadia Vase}
\affiliation{
  \institution{eBay}
  \city{San Jose}
  \country{USA}
}
\author{Shahram Khadivi}
\affiliation{
  \institution{eBay}
  \city{Aachen}
  \country{Germany}
}
\author{Wiktor Walentynowicz}
\affiliation{
  \institution{eBay}
  \city{}
  \country{}
}
\affiliation{
  \institution{Tooploox}
  \city{Wroclaw}
  \country{Poland}
}
\author{Nicola Ueffing}
\affiliation{
  \institution{eBay}
  \city{Aachen}
  \country{Germany}
}
\author{Jan Pawlowski}
\affiliation{
  \institution{eBay}
  \city{}
  \country{}
}
\affiliation{
  \institution{Tooploox}
  \city{Wroclaw}
  \country{Poland}
}
\author{Piotr Szymanski}
\affiliation{
  \institution{eBay}
  \city{}
  \country{}
}
\affiliation{
  \institution{Tooploox}
  \city{Wroclaw}
  \country{Poland}
}
\author{Abraham Bagherjeiran}
\affiliation{
  \institution{eBay}
  \city{San Jose}
  \country{USA}
}

\renewcommand{\shortauthors}{Gniazdowski et al.}

\begin{abstract}
Relevance prediction between user queries and items is a fundamental component of e-commerce search systems. Traditionally, high-quality relevance labels are obtained through human judgment, which is reliable but expensive, slow, and difficult to scale. Recent advances in large language models (LLMs) enable automated relevance judgment at scale, achieving performance comparable to human annotators. However, such systems often rely on large commercial LLMs with complex prompts, resulting in high inference costs and limited throughput.

In this work, we propose a framework that bridges this gap by distilling a high-performing LLM-based relevance judgment system into efficient multimodal student models. We first curate a small, high-quality human-annotated dataset to serve as ground truth. Using this dataset, we develop a reliable \textit{LLM-as-a-judge} pipeline through careful prompt engineering and iterative refinement. This pipeline is then used to generate large-scale relevance labels, which serve as training data for compact multimodal models. Our distilled models achieve performance close to the teacher system while significantly improving latency, cost, and scalability. This approach enables practical deployment of high-quality relevance prediction systems in production environments.
\end{abstract}

\ccsdesc[500]{Information systems~Learning to rank}
\ccsdesc[300]{Information systems~Relevance assessment}
\ccsdesc[100]{Information systems~Retrieval efficiency}
\ccsdesc[100]{Information systems~Multimedia and multimodal retrieval}

\keywords{Multimodal Large Language Models, knowledge distillation, LLM as a Judge, search query relevance prediction}

\maketitle

\section{Introduction}
Relevance prediction is a core task in information retrieval systems and plays a critical role in e-commerce search. Because it directly impacts ranking quality and user experience, accurately estimating the relevance between a user query and candidate items is essential for both evaluation and training of downstream models \cite{ordSim, thomas2023llm}. However, assessing the relevance of query-item pairs is fundamentally a commonsense reasoning task, requiring an understanding of user intent, product attributes, and contextual nuances. As a result, this task can be challenging even for experienced human judges, particularly when queries are ambiguous or item information is incomplete.

Traditionally, relevance labels are obtained through human annotation, where trained judges evaluate query-item pairs based on predefined guidelines. While reliable, this process is costly, time-consuming, and difficult to scale. Large production systems may spend millions annually yet generate only a limited number of labels, creating a bottleneck for model development and evaluation. Moreover, maintaining consistent and high-quality judgments across a wide range of queries and product categories requires significant domain expertise, which is difficult to ensure at scale.

Recent advances in large language models (LLMs) offer a scalable alternative. LLMs have demonstrated strong capabilities in reasoning and evaluation tasks, enabling the \textit{LLM-as-a-judge} paradigm, where models generate relevance judgments given structured instructions and context \cite{faggioli2023perspectives}. Prior work has shown that commercial LLMs can produce relevance labels that align closely with human judgments \cite{soviero2024chatgpt, thomas2023llm}. However, directly deploying LLM-based relevance systems introduces a new challenge. High-performing solutions typically rely on large commercial models combined with long, carefully engineered prompts. While effective, these systems are expensive to run and have limited throughput, making them impractical for large-scale production use.

Taken together, these limitations reveal a fundamental trade-off between accuracy and scalability: human judgments are accurate but expensive and difficult to scale, while LLM-based judgments are scalable but remain computationally costly. To address this challenge, we propose a distillation-based framework for efficient relevance prediction. Prior works suggest that models trained via black-box knowledge distillation—where the token-level output distributions of a large, typically closed-source commercial teacher model are not accessible, can achieve comparable performance \cite{ho2023large}. Our approach first constructs a small, high-quality human-annotated dataset, which is used to develop a reliable \textit{LLM-as-a-judge} pipeline through prompt engineering and iterative refinement. This pipeline is then leveraged to generate large-scale labeled data, which serves as training data for compact multimodal student models. The resulting student models retain much of the teacher system’s performance while significantly improving efficiency, enabling high-throughput, low-latency relevance prediction.

Our contributions can be summarized as follows:
\begin{itemize}
    \item We propose a distillation-based framework that transfers knowledge from a high-performing, prompt-based \textit{LLM-as-a-judge} system into efficient multimodal student models for search relevance prediction.
    
    \item We demonstrate that a commercial, closed-source LLM can be used as a reliable teacher to generate large-scale, high-quality relevance labels, enabling effective black-box distillation in search relevance prediction.
    
    \item We show that the distilled multimodal models achieve performance comparable to the teacher on human-annotated benchmarks, while significantly reducing inference cost and latency.
    
    \item We provide an empirical analysis of model stability, explanation quality, and efficiency trade-offs, demonstrating the suitability of the proposed approach for real-world deployment.
\end{itemize}

\section{Related Work}
In this section, we review prior work on applying LLMs to search relevance prediction, with a focus on \textit{LLM-as-a-judge}, knowledge distillation, and synthetic data generation.

\subsection{LLMs for Relevance Prediction and E-commerce Search}

Large language models (LLMs) and vision-language models (VLMs) have been increasingly adopted in e-commerce applications, including product understanding, recommendation, and search. Prior work has explored adapting general-purpose LLMs to e-commerce domains through large-scale instruction tuning and domain-specific pretraining \cite{ herold2024lilium, herold2025domain, pmlr-v235-peng24c}. Other approaches leverage multimodal architectures to incorporate visual signals for product understanding \cite{xue2023pumgpt, nulli2026adapting, zhao-etal-2024-unifashion}.

In the context of search relevance prediction, both encoder-based and decoder-based models have been studied. Several works employ LLM-based teachers to supervise smaller models via distillation \cite{wang2024improving, zhao2025explainable, shang2025knowledge}, while others propose multi-stage pipelines combining supervised fine-tuning, reasoning, and reinforcement learning \cite{tang2025lref, dong2025taosr1, yang2025taosr}. Despite these advances, most existing approaches rely either on expensive large models at inference time or require complex training pipelines, limiting their scalability in production environments.

\subsection{\textit{LLM-as-a-Judge} and Automated Evaluation}

Recent work has introduced the \textit{LLM-as-a-judge} paradigm, where large language models are used to evaluate outputs based on structured prompts and contextual instructions \cite{faggioli2023perspectives}. Empirical studies demonstrate that LLM-based evaluators can achieve strong agreement with human judgments across a variety of tasks, including dialogue evaluation, reasoning, and information retrieval \cite{soviero2024chatgpt, thomas2023llm}. 

At the same time, prior research highlights that the effectiveness of LLM-based evaluation is highly sensitive to prompt design and reasoning strategies \cite{brown2020fewshot, wei2022cot}. Structured prompting, including detailed instructions, examples, and explicit rules, has been shown to significantly improve evaluation quality. However, deploying such systems in practice often requires complex, carefully engineered prompts and large models, resulting in high computational cost and limited throughput.

\subsection{Knowledge Distillation and Synthetic Data Generation}

Knowledge distillation has been widely used to transfer knowledge from large models to smaller, more efficient ones \cite{ho2023large}. In the context of LLMs, previous works have studied both white-box distillation, where the student has access to the teacher’s raw predictive distributions, and black-box distillation, where only the teacher’s output text is available \cite{gu2024minillm}. Several approaches leverage teacher-generated labels and explanations to improve student model performance \cite{agrawal2025rationale, zhao2025explainable}.

Closely related to our work, recent studies demonstrate the effectiveness of using LLMs to generate large-scale synthetic training data, enabling scalable supervision without human annotation. These approaches show that high-quality synthetic data can significantly improve downstream model performance while reducing reliance on costly human labeling {\cite{abdin2024phi, wang2025distilqwen2, kang2025neurips-distilling}}.

Unlike prior work that either relies on large LLMs at inference time or focuses on distillation without leveraging structured evaluation, our approach unifies \textit{LLM-as-a-judge} evaluation, large-scale synthetic data generation, and multimodal knowledge distillation into a single framework. This enables the use of high-capacity commercial models as teachers while producing efficient models suitable for real-world deployment.

\section{Methodology}

We present the complete study pipeline below, including prompt engineering for the teacher model, automated construction of the student training dataset, and knowledge distillation from the teacher to the student models.

\subsection{Ground Truth Curation}

We define relevance prediction as the task of assigning a graded relevance label to a query-item pair, reflecting how well the item satisfies the user’s intent. Following standard practices in information retrieval, relevance is commonly modeled using graded or ordinal labels \cite{voorhees2002philosophy, ordSim}. In our setting, we adopt a four-level relevance scale: \textit{Bad}, \textit{Fair}, \textit{Good}, and \textit{Excellent}. While these labels provide a high-level abstraction, in practice each grade is governed by detailed annotation guidelines that specify how to handle factors such as exact matches, partial matches, substitutions, bundles, and other edge cases commonly encountered in e-commerce. To ground our system, we construct two human-annotated datasets of query-item pairs: a high-quality dataset annotated by three independent judges per example, and an additional dataset annotated by a single judge to increase coverage (hereafter referred to as the Test 1 and Test 2 datasets). The dataset annotated by three independent judges per example serves as a reference for calibrating the \textit{LLM-as-a-judge} pipeline and evaluating its alignment with human judgments. The distribution of relevance labels for both datasets is provided in Appendix \ref{sec:gt_distribution}.

\subsection{Prompting Strategy} \label{section:prompting_strategy}

The effectiveness of the \textit{LLM-as-a-judge} pipeline critically depends on prompt design, which defines how the model interprets the relevance evaluation task. We adopt a structured prompting strategy that encodes a consistent decision process, including task instructions, definitions of the relevance scale, explicit rules for handling common edge cases, and a set of curated few-shot examples. This design guides the model to reason systematically about query intent and item attributes, reducing ambiguity and improving consistency in predictions. The prompt is iteratively refined using feedback from ground truth annotations and systematic error analysis, where failure cases are analyzed and incorporated as additional rules or examples. This iterative process improves both the accuracy and robustness of the model’s outputs. The resulting prompt enables the model to act as a reliable evaluator, which is essential for the subsequent \textit{LLM-as-a-judge} pipeline used to generate large-scale training data for distillation.

\subsection{LLM-as-a-Judge Pipeline}

We formulate relevance prediction within the \textit{LLM-as-a-judge} paradigm, where a large language model evaluates a query-item pair based on structured instructions and contextual information \cite{faggioli2023perspectives, llm_judge_survey}. Given an input $x$ and context $C$ (prompt), the evaluation output is defined as:
\[
\mathcal{E} \leftarrow \mathcal{P}_{\text{LLM}}(x \oplus C),
\]
where $\mathcal{E}$ denotes the predicted relevance label and an accompanying explanation.

In our setting, the input consists of a query $q$ (user query text) and item metadata $m$, comprising a combination of textual and visual signals that characterize the item listing. The evaluation model $M$ is instantiated as a prompted LLM and produces:
\[
[r, e] \leftarrow M(p, q, m),
\]
where $p$ is a structured prompt, $r$ is the relevance label, and $e$ is a natural-language explanation.

In our framework, the model $M$ serves as a \textbf{teacher model} within the \textit{LLM-as-a-judge} pipeline. It is used to generate large-scale, high-quality supervision data for training downstream student models. For each query-item pair, the teacher produces both a relevance label and an explanation, forming training examples of the form $(q, m, r, e)$.

Both components of the teacher output play an important role. The label provides the prediction target, while the explanation captures the reasoning process underlying the decision. This richer supervision signal improves the ability of the student model to approximate the teacher’s behavior during distillation.

Empirically, we find that a \textit{respond-then-explain} strategy—where the model first predicts the relevance label and then generates an explanation—yields more stable and reliable outputs compared to alternative reasoning strategies, while maintaining high efficiency.

We evaluate multiple commercial and open-source LLMs, including models from the GPT family. Among all evaluated models, \texttt{GPT-4.1} achieves the highest performance, reaching a multi-class relevance accuracy of approximately 83.7\% on the human-annotated benchmark, demonstrating strong alignment with human judgments. Results for other models are provided in Appendix Table~\ref{tab:results_commercial_llms}. 

This strong alignment enables us to use \texttt{GPT-4.1} as a reliable teacher to generate large-scale, high-quality relevance labels and explanations, which are subsequently used to train efficient multimodal student models via knowledge distillation.

\subsection{Distillation to Student Models}

We train a compact multimodal autoregressive model, denoted \(M_{\mathrm{sft}}\), to predict relevance labels and optionally generate explanations:
\[
[r_{\mathrm{sft}}, e_{\mathrm{sft}}] \leftarrow M_{\mathrm{sft}}(q, m).
\]

Our goal is to approximate the behavior of the teacher model \(M\) (\texttt{GPT-4.1}), which produces predictions conditioned on a long prompt \(p\):
\[
[r, e] \leftarrow M(p, q, m).
\]

To enable this distillation, we construct a large-scale dataset using the \textit{LLM-as-a-judge} pipeline. For each example, we form:
\[
X \leftarrow [p_{\mathrm{sft}}, q, m], \quad y \leftarrow [r, e],
\]
where \(p_{\mathrm{sft}}\) is a lightweight prompt suitable for efficient inference (see Appendix~\ref{sec:sft_prompt}), and \((r,e)\) are generated by the teacher model. The resulting dataset \(D = \{(X,y)\}\) captures the teacher’s labeling behavior at scale.

We train \(M_{\mathrm{sft}}\) via supervised finetuning starting from an instruction tuned base model \(M_{\mathrm{it}}\), selected from the \texttt{Gemma-3} \cite{gemmateam2025gemma3technicalreport} family of multimodal large language models. This choice is motivated by native support for text and image inputs, multilingual capabilities, and strong performance across benchmarks. Training is performed using the LLaMA-Factory framework \cite{zheng2024llamafactory}. The student model learns to replicate both relevance labels and explanations, effectively distilling the teacher’s behavior into a compact and efficient architecture suitable for production large-scale deployment.

{For deployment, the student model can operate in a lightweight mode by generating only the relevance label \\$r \in \{\texttt{bad}, \texttt{fair}, \texttt{good}, \texttt{excellent}\}$, significantly reducing inference cost and latency. When required, explanations can also be generated by increasing the output length, without modifying the model architecture.}

\subsection{Datasets}

The training and validation datasets are generated using the \textit{LLM-as-a-judge} pipeline and are balanced across relevance classes and product categories to ensure diversity. In contrast, evaluation is conducted on human-annotated datasets: Test 1 (with 239 samples), which is annotated by three judges and used for prompt engineering for the teacher model, and Test 2 (with 233 samples) annotated by a single judge, completely independent from the prompt creation process. The relevance scores in all datasets are: \textit{Bad, Fair, Good, Excellent}. The explanations are concise, typically consisting of a few sentences describing the key reasoning behind the assigned label.

Model selection is performed using validation data generated by the teacher model, while final evaluation relies on human-labeled data. This setup ensures scalability in training while maintaining reliable evaluation against high-quality human judgments. Additional details on data construction and preprocessing are provided in Appendix \ref{sec:data_details}. Dataset details are provided in Table \ref{tab:data_summary}.

\begin{table}[h!]
    \centering
    \begin{tabular}{lp{5cm}}
        \toprule
        Dataset & Description \\
        \midrule
        Train (483,527) & Training dataset generated with \texttt{GPT-4.1}. Samples are balanced across class labels and categories. \\
        \midrule
        Validation 1 (1,500) & Validation set generated with \texttt{GPT-4.1} and balanced across class labels. This split contains queries that also appear in the training set (of course for different items), resulting in broader category coverage and serving as an in-distribution validation benchmark. \\
        \midrule
        Validation 2 (2,466) & Validation set generated with \texttt{GPT-4.1} and balanced across class labels and category names. Model selection is based on performance on this split. In contrast to Validation 1, it contains queries that do not appear in the training set, making it more challenging despite its narrower category coverage. \\
        \midrule
        Test 1 (239) & Test set annotated independently by three human annotators. This split primarily contains edge cases and particularly challenging examples. This set is used in the process of optimizing prompt for the \textit{LLM-as-a-judge} pipeline. \\
        \midrule
        Test 2 (233) & Test set annotated by a single human annotator. This dataset is completely independent of the prompt engineering process for \textit{LLM-as-a-judge}. \\
        \bottomrule
    \end{tabular}
    \caption{Summary of the datasets used in our experiments, including split sizes, construction protocol, and annotation procedure. Validation 1 and Validation 2 differ in their degree of overlap with the training dataset queries, while Test 1 and Test 2 are human-annotated evaluation sets. Number of samples in each dataset is provided inside brackets.}
    \label{tab:data_summary}
\end{table}

\section{Experiments}

Initial finetuning configs are based on the supervised finetuning configurations of the State of the Art multimodal models: \\\texttt{DeepSeek-VL-7B} (\textit{Stage 3}) \cite{lu2024deepseek}, \texttt{Qwen-VL} (\textit{Stage 3}) \cite{bai2023qwen}, \texttt{NVIDIA-Nemotron-Nano-V2-VL} (\textit{Stage 4}) \cite{deshmukh2025nvidia}, \texttt{Qwen-2.5-VL} (\textit{SFT}) \cite{bai2025qwen3} and \texttt{Gemma-3} \cite{gemmateam2025gemma3technicalreport}. 

Next, we perform iterative hyperparameter tuning for all three \texttt{Gemma-3-IT} variants, namely the 4B, 12B, and 27B versions. For each model, we sequentially optimize a set of training hyperparameters. The hyperparameters considered, in the order in which they are tuned, are listed below. The best model is found based on the smallest loss value on Validation 2 dataset:
\begin{itemize}
    \item number of epochs (from values $\in \{2, 3\}$),
    \item batch size (from values $\in \{256, 512, 1024, 2048, 4096\}$),
    \item weight decay (from values $\in \{0.05, 0.1\}$),
    \item learning rate	(from values $\in \{1e^{-5}, 2e^{-5}, 5e^{-5}, 1e^{-4}\}$).
\end{itemize}
As in the \texttt{Gemma-3-IT} model SFT setup, we freeze the vision encoder in our models. The reason for that is to add regularization to the entire training process. As shown in Table \ref{tab:data_summary}, training and validation datasets are generated using the \texttt{GPT-4.1} teacher model. Test datasets, most of which contain edge cases, are annotated by human annotators. Freezing the vision encoder is intended to introduce a regularization factor to prevent the model from overfitting on synthetic and general data samples.

On this basis, the best sets of hyperparameters are found for finetuning the \texttt{Gemma-3-4B}, \texttt{Gemma-3-12B}, \texttt{Gemma-3-27B} models creating \texttt{Gemma-3-4B-IT-RLV}, \texttt{Gemma-3-12B-IT-RLV}, \\\texttt{Gemma-3-27B-IT-RLV} LLMs trained and evaluated on the datasets described above. The final best hyperparameter values are presented in Appendix Table \ref{tab:hyperparameters}.

\subsection{Optimizer}
AdamW optimizer \cite{loshchilov2018decoupled} is used for training with hyperparameters $\beta_1=0.9$ and $\beta_2=0.999$. Additionally model is trained using cosine learning rate scheduler \cite{loshchilov2017sgdr} with warm-up. It starts training at a very low learning rate value, increases it gradually to a peak over the warm-up fraction (in our setup equal to 0.1), and then decreases it following a cosine-shaped curve. This method helps stabilize very early training steps, when updates of the model weights are noisy, while the smooth cosine decay later allows smaller, gentler updates as the model converges.

\subsection{Hardware}
All training and evaluation experiments are conducted on the same hardware and software setup, using 8 NVIDIA A100 GPUs with CUDA 12.6. Only the throughput and performance experiments, described in the Section \ref{sec: throughput_performance}, are conducted on different hardware—a single NVIDIA H100 GPU. Throughput is measured in FP8, while all models are otherwise run in FP16 precision. The implementation is based on the Llama Factory \cite{zheng2024llamafactory}, Pytorch 2.6.0 \cite{paszke2019pytorch} and Python 3.10, running on Ubuntu 22.04.

\section{Results}
The goal of this task is to achieve performance comparable to the teacher model on the Test 1 and Test 2 datasets, while significantly reducing inference cost and latency relative to the expensive \texttt{GPT-4.1} model. These datasets are annotated by three and one independent human judges, respectively, and contain many challenging examples that require substantial e-commerce expertise. {\color{black}{Test 1 is useful for challenging edge cases but is not a fully untouched benchmark for the teacher pipeline; Test 2 is the cleanest estimate of independent human-evaluation generalization.}} The results for the student models—\texttt{Gemma-3-4B-IT-RLV}, \texttt{Gemma-3-12B-IT-RLV}, \texttt{Gemma-3-27B-IT-RLV}  {\color{black}and encoder-based, text-only baseline model--finetuned on the same data as presented \texttt{Gemma-3-IT-RLV} models, without images and explanations, with classification manner \texttt{modernBERT-large} \cite{warner-etal-2025-smarter} are reported in Table~\ref{tab:results}. The task of the baseline model is to predict the same four-class relevance score as the student models, but without any explanation.} The teacher model operates on lower-quality images (maximum edge length of 300 pixels), whereas the student models are trained and evaluated using higher-quality images (maximum edge length of 900 pixels). For completeness, results obtained using lower-quality images for the student models are provided in Appendix Table~\ref{tab:results_max_image_edge}. Additionally, Appendix Table~\ref{tab:results_dataset_version} analyzes the impact of training data size and diversity on model performance{\color{black}, Appendix Table~\ref{tab:results_explanation} compares training with and without explanations, and Appendix Table~\ref{tab:results_image_usage} compares finetuning with and without images.

The ablation results show that both dataset scale and dataset diversity have a strong positive impact on model quality, as demonstrated in Appendix Section~\ref{section: ablation_number_samples}. Training on the larger and more diverse dataset improves performance across almost all the models and evaluation datasets, with particularly clear gains for underrepresented classes and smaller models. Appendix Section~\ref{section: ablation_image_size} shows that increasing the maximum image edge length from 300 to 900 pixels further improved results, especially for larger models and more difficult cases requiring subtle visual recognition. The results in Appendix Section~\ref{section: explanation_generation} indicate that adding explanations during training improved performance for most model and dataset variants, with the largest benefits observed on edge cases. Section~\ref{section: ablation_image_usage} in Appendix confirms that item images are an important input modality for reliable prediction.}

Notably, image resolution has minimal impact on inference latency for \texttt{Gemma-3} models, as inputs are internally resized to fixed dimensions and presented as 256 visual tokens. 

Overall, the student models achieve performance comparable to the teacher on the Test~1 dataset. On Test 2—an independent dataset that is not used during prompt design for the \textit{LLM-as-a-judge} pipeline—the 12B model achieves even stronger results in terms of both accuracy and F1 score. Performance on validation datasets further indicates that the student models effectively replicate the teacher’s behavior across most examples, demonstrating the effectiveness of the proposed knowledge distillation pipeline.

\begin{table*}[h!]
    \centering
    \begin{tabular}{lcccccccccc}
        \toprule
        & \multicolumn{2}{c}{\texttt{GPT-4.1}} & \multicolumn{2}{c}{\texttt{modernBERT-large-RLV}} & \multicolumn{6}{c}{\texttt{Gemma-3-IT-RLV} \textbf{(our)}} \\
        \cmidrule(lr){6-11} 
        & \multicolumn{2}{c}{Teacher} & \multicolumn{2}{c}{Encoder Baseline} & \multicolumn{2}{c}{\texttt{4B}} & \multicolumn{2}{c}{\texttt{12B}} & \multicolumn{2}{c}{\texttt{27B}} \\
        \cmidrule(lr){2-3} \cmidrule(lr){4-5} \cmidrule(lr){6-7} \cmidrule(lr){8-9} \cmidrule(lr){10-11}
        Dataset & acc [\%] & F1 [\%] & acc [\%] & F1 [\%] & acc [\%] & F1 [\%] & acc [\%] & F1 [\%] & acc [\%] & F1 [\%] \\
        \midrule
        Validation 1 & — & — & 69.93 & 69.80 & 79.33 & 79.22 & \textbf{82.33} & \textbf{82.39} & 81.47 & 81.50 \\
        Validation 2 & — & — & 72.30 & 71.74 & 79.52 & 79.11 & 82.20 & 81.82 & \textbf{83.33} & \textbf{82.96} \\
        Test 1 & \textbf{83.68} & \textbf{81.30} & 74.06 & 61.29 & 80.34 & 76.19 & \underline{81.59} & 77.78 & \underline{81.59} & \underline{78.71} \\
        Test 2 & 64.38 & 56.09 & 57.08 & 47.40 & 63.95 & 56.47 & \textbf{65.24} & \textbf{57.56} & 62.23 & 54.37 \\
        \bottomrule
    \end{tabular}
    \caption{Accuracy and F1 macro obtained for the finetuned models for search relevance classification for the \texttt{Gemma-3-IT-RLV} family models compared to the teacher \texttt{GPT-4.1} model on all evaluation datasets.}
    \label{tab:results}
\end{table*}

\subsection{Per-class Results}
\begin{figure}[h!]
    \centering
    \begin{subfigure}[t]{0.23\textwidth}
        \centering
        \includegraphics[width=\linewidth]{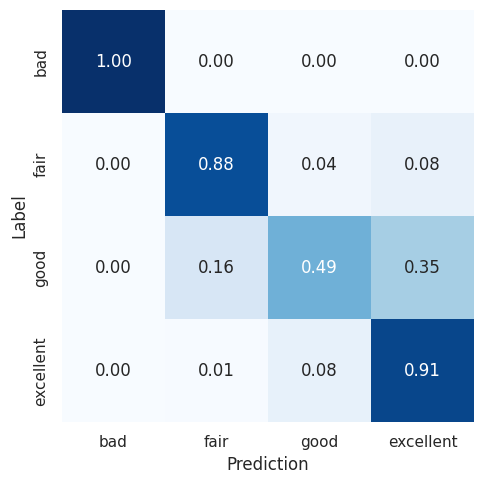}
        \caption{\texttt{GPT-4.1}}
    \end{subfigure}
    \hfill
    \begin{subfigure}[t]{0.23\textwidth}
        \centering
        \includegraphics[width=\linewidth]{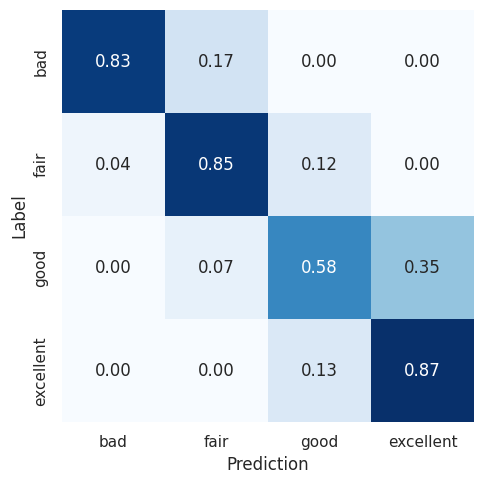}
        \caption{\texttt{Gemma-3-12B-IT-RLV} \textbf{(our)}}
    \end{subfigure}
    \hfill
    \caption{Confusion matrices for teacher and \texttt{Gemma-3-12B-IT-RLV} models on Test 1 dataset.}
    \label{fig:confusion_matrices_test_1}
\end{figure}

\begin{figure}[h!]
    \centering
    \begin{subfigure}[t]{0.23\textwidth}
        \centering
        \includegraphics[width=\linewidth]{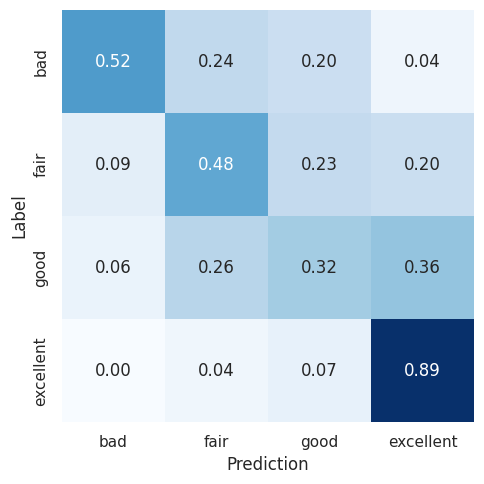}
        \caption{\texttt{GPT-4.1}}
    \end{subfigure}
    \hfill
    \begin{subfigure}[t]{0.23\textwidth}
        \centering
        \includegraphics[width=\linewidth]{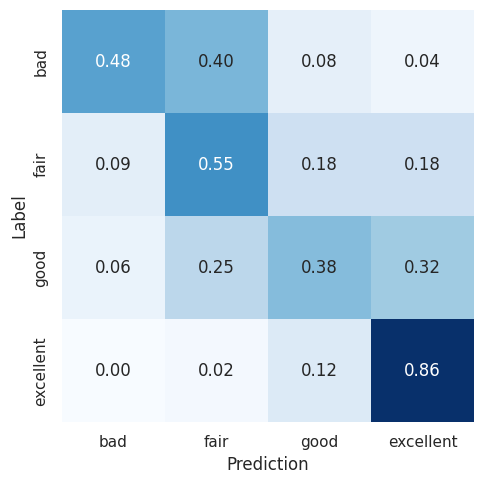}
        \caption{\texttt{Gemma-3-12B-IT-RLV} \textbf{(our)}}
    \end{subfigure}
    \hfill
    \caption{Confusion matrices for teacher and \texttt{Gemma-3-12B-IT-RLV} models on Test 2 dataset.}
    \label{fig:confusion_matrices_test_2}
\end{figure}

The F1 metric values obtained per class for the finetuned models and the teacher's model are presented in Table \ref{tab:results_per_class}. As expected, the middle classes—Fair and Good are more difficult for the models, they usually require more knowledge, and the differences between them are not as obvious as between the extreme classes—Bad and Excellent. This rule is visible for both the teacher model and the finetuned models. Similar conclusions can be drawn from the confusion matrices obtained for all the models on Test 1 and Test 2 datasets (Figures \ref{fig:confusion_matrices_test_1} and \ref{fig:confusion_matrices_test_2})—they are clearly diagonal, with weaker results obtained for the middle classes. 

Based on the confusion matrices for Test 1 dataset, it can also be seen that most of the errors made by all analyzed models are equal to one step of difference between the prediction and the label value which is acceptable. The results for the Test 2 dataset are worse, but this was assumed, as it is a very demanding dataset, requiring extensive e-commerce knowledge and thorough analysis of all input signal features.

Analyzing the results of the models for the extreme and middle classes, it can also be noticed that the finetuned 12B model turned out to be more effective than the teacher model for both Test 1 and Test 2 datasets for middle classes. This is interesting because these samples are the ones that require more extensive e-commerce knowledge and very detailed analysis. 

To sum up, when analyzing the per-class results, it can be said that the student models obtain similar results to the teacher model.

\begin{table*}[h!]
    \centering
    \begin{tabular}{llcccc}
        \toprule
        Dataset & Model & Bad [\%] & Fair [\%] & Good [\%] & Excellent [\%] \\
        \midrule
        \multirow{4}{*}{Test 1} & \texttt{GPT-4.1} & \textbf{100.00} & 80.70 & 53.85 & \textbf{90.63} \\
        \cmidrule(lr){2-6}
        & \texttt{Gemma-3-4B-IT-RLV} \textbf{(our)} & \underline{92.31} & 76.00 & 47.62 & 88.82 \\
        & \texttt{Gemma-3-12B-IT-RLV} \textbf{(our)} & 83.33 & \textbf{84.62} & \textbf{54.35} & 88.82 \\
        & \texttt{Gemma-3-27B-IT-RLV} \textbf{(our)} & \underline{92.31} & 82.35 & 51.16 & \underline{89.02} \\
        \midrule
        \midrule
        \multirow{4}{*}{Test 2} & \texttt{GPT-4.1} & \textbf{57.78} & 47.19 & 36.56 & \textbf{82.85} \\
        \cmidrule(lr){2-6}
        & \texttt{Gemma-3-4B-IT-RLV} \textbf{(our)} & 53.33 & 50.00 & 41.18 & 81.39 \\
        & \texttt{Gemma-3-12B-IT-RLV} \textbf{(our)} & \underline{54.55} & \textbf{51.61} & \textbf{41.67} & \underline{82.40} \\
        & \texttt{Gemma-3-27B-IT-RLV} \textbf{(our)} & 51.16 & 47.83 & 36.89 & 81.58 \\
        \bottomrule
    \end{tabular}
    \caption{F1 metric values for search relevance classification for the \texttt{Gemma-3-IT-RLV} family models compared to the teacher \texttt{GPT-4.1} model per class on Test 1 and Test 2 datasets.}
    \label{tab:results_per_class}
\end{table*}

\subsection{Stability}
Results for stability experiment for \texttt{Gemma-3-4B-IT-RLV}, \\ \texttt{Gemma-3-12B-IT-RLV}, \texttt{Gemma-3-27B-IT-RLV} models are presented in Table \ref{tab:results_stability}. For this experiment all finetuned models are evaluated 20 times on the Test 1 dataset to verify prediction stability. The table contains the mean and standard deviation values of the obtained results. With 239 samples in Test 1 dataset, one sample is 0.42\% of the whole dataset. The variation between runs is very small: the standard deviation is lower than the share represented by a single example in the full Test 1 dataset. It can therefore be said that the predictions of all three models are stable across all the runs.
\begin{table}[h!]
    \centering
    \begin{tabular}{lc}
         \toprule
         Model & Accuracy \\
         \midrule
         \texttt{Gemma-3-4B-IT-RLV} & (80.57 $\pm$ 0.21)\% \\
         \texttt{Gemma-3-12B-IT-RLV} & \textbf{(81.63 $\pm$ 0.29)\%} \\
         \texttt{Gemma-3-27B-IT-RLV} & (81.38 $\pm$ 0.25)\% \\
         \bottomrule
    \end{tabular}
    \caption{Mean accuracy and standard deviation values for all finetuned models for 20 independent model runs for search relevance classification.}
    \label{tab:results_stability}
\end{table}

\subsection{Explanation}

The quality of the generated explanations for \texttt{Gemma-3-4B-IT-RLV}, \texttt{Gemma-3-12B-IT-RLV}, \texttt{Gemma-3-27B-IT-RLV} models are also examined. The chrf \cite{popovic-2015-chrf} metric and the BERTScore F1 \cite{Zhang*2020BERTScore:} are calculated for both lower-case predictions and labels and presented in Table \ref{tab:results_explanation_metrics}. The chrf metric measures how similar the characters in the prediction and label are, while the BERTScore metric measures how similar the two texts are in meaning. High values of both metrics—especially BERTScore, which measures the semantic similarity of the texts, indicate that the quality of the generated explanations is high for all finetuned models. Please note here that the explanations are only in the validation datasets. The test datasets contain only relevance scores.

\begin{table}[h!]
    \centering
    \begin{tabular}{llcc}
    \toprule
    Dataset & Model & chrf [\%] & BS F1 [\%] \\
    \midrule
        \multirow{3}{*}{Val 1} 
        & \texttt{Gemma-3-4B-IT-RLV} & 67.15 & 93.78 \\ 
        & \texttt{Gemma-3-12B-IT-RLV}  & 67.97 & 93.89 \\ 
        & \texttt{Gemma-3-27B-IT-RLV} & \textbf{68.42} & \textbf{93.98} \\ 
    \midrule
        \multirow{3}{*}{Val 2} 
        & \texttt{Gemma-3-4B-IT-RLV} & 66.22 & 93.45 \\ 
        & \texttt{Gemma-3-12B-IT-RLV} & 67.48 & 93.70 \\ 
        & \texttt{Gemma-3-27B-IT-RLV} & \textbf{67.68} & \textbf{93.79} \\ 
    \bottomrule
    \end{tabular}
    \caption{Models explanation quality—chrf and BERTScore F1 mean metric values (marked as BS F1 in Table) for all finetuned models from the \texttt{Gemma-3-IT-RLV} family calculated on Validation 1 (Val 1) and Validation 2 (Val 2) datasets.}
    \label{tab:results_explanation_metrics}
\end{table}

\subsection{Throughput and Performance} \label{sec: throughput_performance}
Throughput and accuracy results for \texttt{Gemma-3-12B-IT-RLV} in FP8 and FP16 are reported in Table \ref{tab:results_throughput_performance}. All measurements are performed on a single NVIDIA H100 GPU, which supports FP8 computation. The results show that FP8-quantized models maintain comparable accuracy to their FP16 counterparts, while providing a substantial increase in maximum throughput. Given the high output quality of \texttt{Gemma-3-12B-IT-RLV} relative to \texttt{Gemma-3-27B-IT-RLV} (see Tables \ref{tab:results}, \ref{tab:results_per_class}, \ref{tab:results_stability}), together with its high maximum throughput and strong quantization performance (see Table \ref{tab:results_throughput_performance}), we select \\\texttt{Gemma-3-12B-IT-RLV-FP8} as the model that best represents the Pareto frontier sweet spot. In particular, it offers the best trade-off between metric performance and throughput, achieving strong evaluation results while maintaining very high serving efficiency. {\color{black}Appendix Table \ref{tab:cost_comparison} provides a comparison of the cost of 1 million input tokens and 1 million output tokens for the teacher and \texttt{Gemma-3-12B-IT-RLV-FP8} models. The results indicate that the student models are substantially more cost-effective in both input- and output-token costs relative to GPT-4.1.}

\begin{table}[h!]
    \centering
    \begin{tabular}{lcc}
         \toprule
         Dataset & \multicolumn{2}{c}{\texttt{Gemma-3-12B-IT-RLV}} \\
         & FP16 & FP8 \\  
         \midrule
         Validation 1 [\%] & 82.33 & 82.00 \textcolor{red}{[-0.33]} \\
         Validation 2 [\%] & 82.20 & 81.87 \textcolor{red}{[-0.33]} \\
         Test 1 [\%] & 81.59 & 81.17 \textcolor{red}{[-0.42]} \\
         Test 2 [\%] & 65.24 & 65.67 \textcolor{green}{[+0.43]} \\
         TP w/o explan. [req/s] & 194.52 & 271.54 \textcolor{green}{[+77.02]} \\
         TP w/ explan. [req/s] & 49.57  & 55.98 \textcolor{green}{[+6.41]} \\
         \bottomrule
    \end{tabular}
    \caption{Accuracy [\%] and throughput for the \texttt{Gemma-3-12B-IT-RLV} models quantized to FP16 and FP8. Throughput (marked in Table as TP) is reported in requests per second [req/s]. Experiments are conducted on a single NVIDIA H100 GPU.}
    \label{tab:results_throughput_performance}
\end{table}

\section{Conclusion}
In this work, we presented a scalable framework for search relevance prediction that bridges the gap between high-quality but expensive human and LLM-based judgment systems and efficient production-ready models. By leveraging a carefully designed \textit{LLM-as-a-judge} pipeline, we generate large-scale, high-quality relevance labels that enable effective knowledge distillation into compact multimodal student models. Experimental results demonstrate that the distilled \texttt{Gemma-3-IT-RLV} models achieve performance comparable to the \texttt{GPT-4.1} teacher on human-annotated benchmarks, while significantly reducing inference cost and latency. {\color{black}\\\texttt{Gemma-3-12B-IT-RLV-FP8} is the best Pareto point because it is near-teacher quality, stronger than the encoder baseline, competitive with or better than the 27B model variant on the independent human Test 2 dataset, and much faster and cheaper in FP8.} The student models also exhibit strong generalization, stability across runs, and high-quality explanation generation, confirming that they successfully capture the behavior of the teacher system. Overall, our results show that it is possible to retain the accuracy of large, prompt-based LLM systems while enabling efficient, high-throughput relevance prediction suitable for real-world deployment. This work highlights the potential of combining \textit{LLM-as-a-judge} systems with knowledge distillation to build scalable and practical relevance modeling pipelines for large-scale search applications.

\bibliographystyle{ACM-Reference-Format}
\bibliography{references}

\appendix
\section{Test Dataset Distributions} \label{sec:gt_distribution}

We construct two human-annotated datasets to evaluate the alignment between LLM-generated judgments and human relevance assessments. Both datasets use the same four-level relevance scale: \textit{Bad}, \textit{Fair}, \textit{Good}, and \textit{Excellent}.

\subsection{Test 1 Dataset}

The first dataset consists of 239 query-item pairs, each independently annotated by three human judges. The final label for each pair is determined based on the annotated judgments, allowing for more robust and reliable ground truth through agreement across multiple annotators. Table \ref{tab:gt_multi} shows the distribution of relevance labels for this dataset.

\begin{table}[h]
\centering
\begin{tabular}{lcccc}
\toprule
Label & Bad & Fair & Good & Excellent \\
\midrule
Count & 6 & 26 & 43 & 164 \\
\bottomrule
\end{tabular}
\caption{Label distribution for the multi-annotator dataset (239 samples, 3 judges per sample).}
\label{tab:gt_multi}
\end{table}

\subsection{Test 2 Dataset}

The second dataset consists of 233 query-item pairs, each annotated by a single human judge. While this dataset provides additional coverage, it may exhibit higher variability due to the absence of inter-annotator agreement. Table \ref{tab:gt_single} shows the distribution of relevance labels for this dataset.

\begin{table}[h]
\centering
\begin{tabular}{lcccc}
\toprule
Label & Bad & Fair & Good & Excellent \\
\midrule
Count & 25 & 44 & 53 & 111 \\
\bottomrule
\end{tabular}
\caption{Label distribution for the single-annotator dataset (233 samples, 1 judge per sample).}
\label{tab:gt_single}
\end{table}

\section{Data Details} \label{sec:data_details}
Records are preprocessed by removing duplicates, dropping entries with missing mandatory fields (user query, model prediction), filtering to English-language texts only using FastText language detection model \cite{joulin-etal-2017-bag, joulin2016fasttext}, and enforcing minimum/maximum character length thresholds.

After a detailed analysis of the problem, we identify a set of special-case examples that are excluded from the downsampling procedure and retained in the dataset. Specifically, these are data points containing predefined keywords and are therefore preserved in the training or validation splits. The presence of such keywords suggests that a sample may correspond to a challenging edge case that is important for the task, motivating its inclusion despite downsampling. The rest of the data is downsampled. To obtain a relatively balanced dataset, weighted sampling is used, in which the final weight value $w$ is determined from the formula: $w \leftarrow w_{label} + 0.1 * w_{category}$. The sampling weights $w_{label}$ and $w_{category}$ are calculated for respectively: label generated with \texttt{GPT-4.1} (relevance score) and item category with \texttt{compute\_sample\_weight} function from Scikit-learn library \cite{Scikit-learn}.

To prevent data leakage, training candidates are also checked against all evaluation sets using both exact matching and fuzzy logic matching using thefuzz Python library \cite{seatgeek_thefuzz_2023}.

The entire process described above results in the creation of Training and Validation 2 datasets. Because data creation is a long and iterative process, we also have Validation 1 data, which is created in a similar way at initial stages of the study. These datasets are not concatenated because Validation 1 contains user queries that are present in the training data (only for other items). Validation 2 on the other hand does not contain queries that are present in the training dataset—later in the project, we have more data samples available. As a result, Validation 1 dataset has 1500 balanced samples, it contains queries that are in training data, but because of that, the categories of the datapoints are more various. On the other hand Validation 2 dataset contains 2466 balanced samples and contains queries that are not present in the training dataset. Because of that categories of the datapoints are less diverse (item categories are inextricably linked to user queries). However, the created dataset is more difficult, as the queries from Validation 2 that determine the user's needs, do not appear in the Training dataset. For this reason, it is decided to separate both sets and provide two metrics for each of them. Final Training dataset contains 483,527 samples. 

{\color{black}{The resulting datasets span a broad and heterogeneous set of item categories. The most common meta-categories include: collectibles, fashion, electronics, and vehicle parts/accessories. Additional substantial groups correspond to sports equipment, toys, home goods, media, health and beauty, jewellery, and business/industrial supplies, indicating that the dataset captures a diverse cross-section of marketplace inventory.
}}

The Test datasets are also created iteratively and, unlike the validation data, are annotated by human annotators. The cost of obtaining such annotations is enormous, with 239 samples obtained for Test 1 dataset and 233 samples for Test 2 dataset.

{\color{black}{
\section{Results for Other Large Language Models}
Table \ref{tab:results_commercial_llms} summarizes the results for the commercial LLMs: \texttt{GPT-5}, \texttt{GPT-4.1}, \texttt{GPT-4o}, \texttt{GPT-4.1-mini}, and \texttt{GPT-4.1-nano} and open-source LLMs: \texttt{Gemma-3-27B-IT} and \texttt{Qwen3-VL-32B-Instruct} \cite{bai2025qwen3} prompted with the prepared prompt described in Section~\ref{section:prompting_strategy}.
}}
\begin{table}[h!]
    \centering
    \begin{tabular}{ccc}
    \toprule
    Weights & Model & Test 1 [\%] \\
    \midrule
    \multirow{5}{*}{closed-source} & \texttt{GPT-5} & 80.33 \\
    & \texttt{GPT-4.1} & \textbf{83.68} \\
    & \texttt{GPT-4o} & 79.92 \\
    & \texttt{GPT-4.1-mini} & 74.06 \\
    & \texttt{GPT-4.1-nano} & 55.65 \\
    \midrule
    \multirow{2}{*}{open-source} & \texttt{Gemma-3-27B-IT} & {\color{black}{68.20}} \\
    & \texttt{Qwen3-VL-32B-Instruct} & {\color{black}{72.38}} \\
    \bottomrule
    \end{tabular}
    \caption{Accuracy scores [\%] on the Test 1 dataset for the commercial and open-source LLMs. All models are prompted using the prompt described in Section~\ref{section:prompting_strategy}.}
    \label{tab:results_commercial_llms}
\end{table}

\section{Supervised Finetuning Prompt} \label{sec:sft_prompt}
{\color{black}{The prompt used to finetune all \texttt{Gemma-3-IT-RLV} models is shown below. In the production setting, this prompt can be cached, and only the input item data and user query are appended.}}
\begin{promptbox}[SFT prompt]
Based on the given query, item metadata, determine how relevant the item is to the query. Return a value specifying the level of relevance from the list: ['bad', 'fair', 'good', 'excellent'] and explanation.
<item metadata>
<user query>
\end{promptbox}

\section{\texttt{Gemma-3-12B-IT-RLV} vs \texttt{Gemma-3-27B-IT-RLV} Explanation Quality}
Furthermore chrf and BERTScore metrics are calculated between the 12B and 27B models, to determine the degree of explanation quality degradation for the smaller model. Lower-case predictions of \texttt{Gemma-3-27B-IT-RLV} are set as labels and lower-case predictions of \texttt{Gemma-3-12B-IT-RLV} as predictions. For this task, \texttt{Qwen3-32B} \cite{qwen3technicalreport} is also used (a totally different model is used on purpose to avoid introducing bias) as a \textit{LLM-as-a-judge} to assess whether the given texts are same, similar, or completely different in meaning. The LLM prompt for this task can be found below.

Based on the analysis of the results comparing the explanation quality of 12B and 27B models (Table \ref{tab:results_explanation_12B_vs_27B}), the quality of the explanations provided by the smaller model is comparable to those predicted by the larger one. Most of the explanations are assessed by \textit{LLM-as-a-judge} as the same or similar (2\%—4\% of the explanations across all the evaluation datasets are judged as different). The similarity of both explanations across all datasets is also evidenced by the high mean BERTScore F1 metric values ($\approx95\%$ across all datasets).

\begin{promptbox}[Prompt for an \textit{LLM-as-a-judge} text similarity task]
    Task: Compare Text A and Text B and output only one word: same, similar, or different.
    
    Criteria:
    * same: identical meaning or paraphrase; differences are stylistic/trivial,
    * similar: same core idea/intent with minor additions/omissions,
    * different: conflicting meanings, topics, claims, or intents.
    
    Consider negation, scope, time, quantities, and entities. Ignore wording, order, and formatting. Do not return explanation.
    Text A: <text a>
    Text B: <text b>
\end{promptbox}

\begin{table}[h!]
    \centering
    \begin{tabular}{lccccc}
    \toprule
        \multirow{2}{*}{Dataset} & \multirow{2}{*}{chrf} & \multirow{2}{*}{BS F1} & \multicolumn{3}{c}{LLM judge} \\
        \cmidrule(lr){4-6}
        & & & same & similar & different \\
        \midrule
        Val 1 & 75.07 & 95.25 & 875 \textcolor{green}{(58.3\%)} & 590 \textcolor{gray}{(39.3\%)} & 35 \textcolor{red}{(2.3\%)} \\
        Val 2 & 74.81 & 95.06 & 1532 \textcolor{green}{(62.1\%)} & 852 \textcolor{gray}{(34.6\%)} & 82 \textcolor{red}{(3.3\%)} \\
        Test 1 & 75.27 & 95.17 & 142 \textcolor{green}{(59.4\%)} & 91 \textcolor{gray}{(38.1\%)} & 6 \textcolor{red}{(2.5\%)} \\
        Test 2 & 73.27 & 94.79 & 125 \textcolor{green}{(53.7\%)} & 99 \textcolor{gray}{(42.5\%)} & 9 \textcolor{red}{(3.9\%)} \\
    \bottomrule
    \end{tabular}
    \caption{Models explanation quality—chrf, BERTScore F1 mean metric (marked in Table as BS F1) values and \textit{LLM-as-a-judge} results. Here the lower-case \texttt{Gemma-3-27B-IT-RLV} predictions are set as labels and lower-case predictions of \texttt{Gemma-3-12B-IT-RLV} as predictions on Validation 1 and Validation 2 datasets (marked as Val 1 and Val 2 in Table) and Test 1 and Test 2 datasets.}
    \label{tab:results_explanation_12B_vs_27B}
\end{table}

\section{The Best Hyperparameter Sets}
The best found sets of hyperparameters for all finetuned models are presented in Table \ref{tab:hyperparameters}.
\begin{table}[h!]
    \centering
    \begin{tabular}{lccc}
        \toprule
        \multirow{2}{*}{Hyperparameter} & \multicolumn{3}{c}{\texttt{Gemma-3-IT-RLV}} \\
        \cmidrule(lr){2-4} 
        & \texttt{4B} & \texttt{12B} & \texttt{27B} \\
        \midrule
        Learning rate & $2e^{-5}$ & $2e^{-5}$ & $2e^{-5}$ \\
        LR scheduler & cosine & cosine & cosine \\
        Weight decay & 0.1 & 0.1 & 0.1 \\
        Gradient clip & 1.0 & 1.0 & 1.0 \\
        \multirow{2}{*}{Optimizer} & AdamW & AdamW & AdamW \\
        & $(0.9,0.999)$ & $(0.9,0.999)$ & $(0.9,0.999)$ \\
        Warm-up fraction & 0.1 & 0.1 & 0.1 \\
        Epochs & 2 & 2 & 2 \\
        Batch size & 512 & 1024 & 2048 \\
        Sequence length & 1024 & 1024 & 1024 \\
        Freeze vision enc & Yes & Yes & Yes \\
        Freeze mm projector & No & No & No \\
        \bottomrule
    \end{tabular}
    \caption{The best hyperparameter sets found for all analyzed models. Hyperparameters for AdamW optimizer $\beta_1$ and $\beta_2$ are presented in the brackets.}
    \label{tab:hyperparameters}
\end{table}

\section{Ablations}
The influence of the number and variety of samples and image sizes in the training set is investigated. The results are presented in sections \ref{section: ablation_number_samples} and \ref{section: ablation_image_size}

\begin{table*}[h!]
    \centering
    \begin{tabular}{lccccccc}
    \toprule
    Model name & \multirow{2}{*}{Metric} & \multicolumn{2}{c}{\texttt{Gemma-3-4B-IT-RLV}} & \multicolumn{2}{c}{\texttt{Gemma-3-12B-IT-RLV}} & \multicolumn{2}{c}{\texttt{Gemma-3-27B-IT-RLV}} \\
    \cmidrule(lr){3-4} \cmidrule(lr){5-6} \cmidrule(lr){7-8}
    Train dataset version &  & v1 & v2 & v1 & v2 & v1 & v2 \\
    \toprule
    Validation 1 & acc [\%] & 78.00 & 79.00 \textcolor{green}{[+1.00$\uparrow$]} & 80.13 & 81.27 \textcolor{green}{[+1.14$\uparrow$]} & 80.67 & 81.20 \textcolor{green}{[+0.53$\uparrow$]} \\
    \multirow{2}{*}{Test 1} & acc [\%] & 76.15 & 79.50 \textcolor{green}{[+3.35$\uparrow$]} & 75.73 & 79.92 \textcolor{green}{[+4.19$\uparrow$]} & 78.24 & 79.50 \textcolor{green}{[+1.26$\uparrow$]} \\
    & F1 [\%] & 68.92 & 71.03 \textcolor{green}{[+2.11$\uparrow$]} & 64.20 & 76.96 \textcolor{green}{[+12.76$\uparrow$]} & 69.85 & 76.82 \textcolor{green}{[+6.97$\uparrow$]} \\
    Test 2 & acc [\%] & 61.80 & 61.80 \textcolor{gray}{[-]} & 65.24 & 62.66 \textcolor{red}{[-2.58$\downarrow$]} & 63.09 & 64.38 \textcolor{green}{[+1.29$\uparrow$]} \\
    \bottomrule
    \end{tabular}
    \caption{The influence of the dataset version (number of datapoints and their diversity) on the search relevance classification quality.}
    \label{tab:results_dataset_version}
\end{table*}

\begin{table*}[h!]
    \centering
    \begin{tabular}{lccccccc}
    \toprule
    Model name & \multirow{2}{*}{Metric} & \multicolumn{2}{c}{\texttt{Gemma-3-4B-IT-RLV}} & \multicolumn{2}{c}{\texttt{Gemma-3-12B-IT-RLV}} & \multicolumn{2}{c}{\texttt{Gemma-3-27B-IT-RLV}} \\
    \cmidrule(lr){3-4} \cmidrule(lr){5-6} \cmidrule(lr){7-8}
    Max image edge length &  & 300 px & 900 px & 300 px & 900 px & 300 px & 900 px \\
    \midrule
    Validation 1 & acc [\%] & 79.00 & 79.33 \textcolor{green}{[+0.33$\uparrow$]} & 81.27 & 82.33 \textcolor{green}{[+1.06$\uparrow$]} & 81.20 & 81.47 \textcolor{green}{[+0.27$\uparrow$]} \\
    Validation 2 & acc [\%] & 79.40 & 79.52 \textcolor{green}{[+0.12$\uparrow$]} & 81.96 & 82.20 \textcolor{green}{[+0.24$\uparrow$]} & 81.83 & 83.33 \textcolor{green}{[+1.50$\uparrow$]} \\
    \multirow{2}{*}{Test 1} & acc [\%] & 79.50 & 80.34 \textcolor{green}{[+0.84$\uparrow$]} & 79.92 & 81.59 \textcolor{green}{[+1.67$\uparrow$]} & 79.50 & 81.59 \textcolor{green}{[+2.09$\uparrow$]} \\
    & F1 [\%] & 71.03 & 76.19 \textcolor{green}{[+5.16$\uparrow$]} & 76.96 & 77.78 \textcolor{green}{[+0.82$\uparrow$]} & 76.82 & 78.71 \textcolor{green}{[+1.89$\uparrow$]} \\
    Test 2 & acc [\%] & 61.80 & 63.95 \textcolor{green}{[+2.15$\uparrow$]} & 62.66 & 65.24 \textcolor{green}{[+2.58$\uparrow$]} & 64.38 & 62.23 \textcolor{red}{[-2.15$\downarrow$]} \\
    \bottomrule
    \end{tabular}
    \caption{The influence of the dataset's maximum image edge length (300 pixels and 900 pixels) on the search relevance classification quality.}
    \label{tab:results_max_image_edge}
\end{table*}

\begin{table*}[h!]
    \centering
    \begin{tabular}{lccccccc}
    \toprule
    Model name & \multirow{2}{*}{Metric} & \multicolumn{2}{c}{\texttt{Gemma-3-4B-IT-RLV}} & \multicolumn{2}{c}{\texttt{Gemma-3-12B-IT-RLV}} & \multicolumn{2}{c}{\texttt{Gemma-3-27B-IT-RLV}} \\
    \cmidrule(lr){3-4} \cmidrule(lr){5-6} \cmidrule(lr){7-8}
    Explanation generation &  & w/o & w/ & w/o & w/ & w/o & w/ \\
    \midrule
    Validation 1 & acc [\%] & 78.20 & 79.33 {\color{green}{[+1.13$\uparrow$]}} & 78.67 & 82.33 {\color{green}{[+3.66$\uparrow$}]} & 80.60 & 81.47 {\color{green}{[+0.87$\uparrow$}]} \\
    Validation 2 & acc [\%] & 77.45 & 79.52 {\color{green}{[+2.07$\uparrow$]}} & 80.45 & 82.20 {\color{green}{[+1.75$\uparrow$]}} & 80.82 & 83.33 {\color{green}{[+2.51$\uparrow$}]}\\
    \multirow{2}{*}{Test 1} & acc [\%] & 79.08 & 80.34 {\color{green}{[+1.26$\uparrow$]}} & 80.33 & 81.59 {\color{green}{[+1.26$\uparrow$]}} & 80.33 & 81.59 {\color{green}{[+1.26$\uparrow$}]} \\
    & F1 [\%] & 68.32 & 76.19 {\color{green}{[+7.87$\uparrow$]}} & 72.62 & 77.78 {\color{green}{[+5.16$\uparrow$]}} & 70.47 & 78.71 {\color{green}{[+8.24$\uparrow$}]} \\
    Test 2 & acc [\%] & 60.09 & 63.95 {\color{green}{[+3.86$\uparrow$]}} & 65.24 & 65.24 {\color{gray}{[-]}} & 63.95 & 62.23 {\color{red}{[-1.72$\downarrow$}]} \\
    \bottomrule
    \end{tabular}
    \caption{{\color{black}{The influence of finetuning the models on the examples without and with explanations on the search relevance classification quality. Models trained without explanations are trained only on the relevance scores.}}}
    \label{tab:results_explanation}
\end{table*}

\begin{table*}[h!]
    \centering
    \begin{tabular}{lccccccc}
    \toprule
    Model name & \multirow{2}{*}{Metric} & \multicolumn{2}{c}{\texttt{Gemma-3-4B-IT-RLV}} & \multicolumn{2}{c}{\texttt{Gemma-3-12B-IT-RLV}} & \multicolumn{2}{c}{\texttt{Gemma-3-27B-IT-RLV}} \\
    \cmidrule(lr){3-4} \cmidrule(lr){5-6} \cmidrule(lr){7-8}
    Image usage &  & w/o & w/ & w/o & w/ & w/o & w/ \\
    \midrule
    Validation 1 & acc [\%] & 71.80 & 78.20 {\color{green}{[+6.4$\uparrow$}]} & 77.60 & 78.67 {\color{green}{[+1.07$\uparrow$]}} & 79.13 & 80.60 {\color{green}{[+1.47$\uparrow$}]} \\
    Validation 2 & acc [\%] & 74.17 & 77.45 {\color{green}{[+3.28$\uparrow$}]} & 78.26 & 80.45 {\color{green}{[+2.19$\uparrow$]}} & 79.76 & 80.82 {\color{green}{[+1.06$\uparrow$}]} \\
    \multirow{2}{*}{Test 1} & acc [\%] & 72.38 & 79.08 {\color{green}{[+6.7$\uparrow$}]} & 78.66 & 80.33 {\color{green}{[+1.67$\uparrow$]}} & 77.41 & 80.33 {\color{green}{[+2.92$\uparrow$}]} \\
    & F1 [\%] & 62.24 & 68.32 {\color{green}{[+6.08$\uparrow$}]} & 65.33 & 72.62 {\color{green}{[+7.29$\uparrow$]}} & 70.09 & 70.47 {\color{green}{[+0.38$\uparrow$}]} \\
    Test 2 & acc [\%] & 61.37 & 60.09 {\color{red}{[-1.28$\downarrow$}]} & 63.09 & 65.24 {\color{green}{[+2.15$\uparrow$]}} & 61.80 & 63.95 {\color{green}{[+2.15$\uparrow$}]} \\
    \bottomrule
    \end{tabular}
    \caption{\color{black}{The influence of finetuning the models without and with item images on the search relevance classification quality. Since the explanation contains information from the item images, both models are trained without explanations.}}
    \label{tab:results_image_usage}
\end{table*}

\subsection{Number of Data Samples} \label{section: ablation_number_samples}
Because the datapoints are collected iteratively from the production processes, the impact of number of samples on the final model quality can be easily compared. Table \ref{tab:results_dataset_version} shows the impact of sample count and sample diversity on the final model quality. Training set version 2 (v2) is described in Table \ref{tab:data_summary}. Training dataset version 1 (v1) is created from less diverse data, with 52,456 samples. The following results are obtained for all models and the maximum edge length of the images in both datasets is 300 pixels for fair comparison. New hyperparameter sets are also found for all models trained on the smaller dataset.

Using approximately nine times more training data in Train 2 dataset than in Train 1 dataset, together with the inclusion of a broader range of special cases and diverse examples across more categories, led to a consistent improvement in the performance of all the trained models across all evaluation datasets. The only exception is \texttt{Gemma-3-12B-IT-RLV} on Test 2 dataset. Models trained on the larger dataset (v2) improve accuracy by 3.35 p.p., 4.19 p.p. and 1.26 p.p. over the models trained on dataset v1 on Test 1 dataset. The difference in the prediction quality for Test 1 dataset of the models is especially visible for the F1 macro metric values, which take into account the prediction quality for underrepresented classes (Test 1 dataset is unbalanced). As expected, using a larger number of more diverse examples improves the model's performance on underrepresented classes. Smaller models (4B and 12B) generally benefit more from v2 dataset than 27B model. This is probably because the 27B model starts with broader prior knowledge. It is also worth noting that the 12B and 27B models trained on v2 dataset achieved similar results across all datasets, so the difference resulting from the different initial level of the models knowledge is minimized after training the models on larger dataset. Interestingly, the transition from Train dataset 1 to Train dataset 2 often yields larger gains than increasing model size, for example when comparing the 4B v2 model against the 12B v1 model, and the 12B v2 model against the 27B v1 model.

\subsection{Image Size} \label{section: ablation_image_size}
Then the impact of image size is compared on the final model quality. The results are presented in Table \ref{tab:results_max_image_edge}. The results from the previous experiment for all the models trained on Training 2 dataset with images with a maximum edge length of 300 pixels (low quality) are compared with the best models trained on the same examples with images having maximum edge length of 900 pixels (high quality).

On Test 1 dataset usage of bigger images (900 pixels instead of 300 pixels) improves accuracy by 0.84 p.p., 1.67 p.p. and 2.09 p.p. for (respectively) 4B, 12B and 27B models. The bigger the model is, the greater improvement is achieved for accuracy on both Validation 2 and Test 1 datasets. Also, the results for 12B and 27B models remain similar, while those for the 4B model remain weaker. Higher resolution seems especially helpful for more difficult predictions. The especially strong F1 gain for the 4B model on Test 1 and accuracy metric on Test 2 for both 4B and 12B models suggests that larger images may improve recognition of subtle visual patterns and special cases.

\subsection{Explanation Generation} \label{section: explanation_generation}
{\color{black}A comparison of the results obtained on the evaluation data for the created models finetuned on the datasets without and with explanations is presented in Table \ref{tab:results_explanation}. The benefits of training the models on examples with explanations are particularly pronounced on the edge cases from Test 1. This is plausible when training on explanations enables the model to better capture subtle differences in the input data. More broadly, finetuning with explanations improves performance for most of the models and dataset variants.}

\subsection{Image usage} \label{section: ablation_image_usage}
{\color{black}A comparison of the results obtained for the models finetuned without and with item images is presented in Table \ref{tab:results_image_usage}. To ensure a fair comparison that isolates a single missing input signal, we train both models without explanations. This design choice is important because, in most cases, the explanation itself contains information derived from the item images. Excluding explanations from training prevents the model from learning to hallucinate visual information that is unavailable at inference time.

This study shows that images are an important modality that usually improves performance. The benefit is most pronounced for the smallest model, with 4B parameters, which achieves the largest performance gains from incorporating images across all datasets, except Test 2, relative to the 12B and 27B models.}

\section{Cost comparison}\label{section:cost_comparison}
{\color{black}
A comparison of the cost per 1 million input and 1 million output tokens is provided in Table~\ref{tab:cost_comparison}. Token cost for \texttt{GPT-4.1} is obtained from the OpenAI pricing page \cite{openai_pricing_2026}. Assuming 120 input text tokens, 256 input image tokens, and 130 output text tokens per request in the explanation-generation setup, and using the measured \texttt{Gemma-3-12B-IT-RLV-FP8} throughput on a single H100 as a proxy, the estimated pure compute cost without any prompt caching is about \$11.86—\$19.80 per 1M requests when using an \textbf{external} GPU provider \cite{runpod_pricing_2026,lambda_pricing_2026}. Per 1 million tokens, this corresponds to about \$0.0234-\$0.0391 for both 1 million input and output tokens, assuming maximum GPU utilization. In the label-only setup without explanation generation, the estimated cost is about \$2.45—\$4.08 per 1M requests, and \$0.0065-\$0.0108 for both 1 million input and 1 million output tokens. When using dedicated in-house GPUs, these savings are naturally even more substantial.}
\newpage
\begin{table}[h!]
    \centering
    \begin{tabular}{cccc}
         \toprule
         Model name & \texttt{GPT-4.1} & \multicolumn{2}{c}{\texttt{Gemma-3-12B-IT-RLV-FP8}} \\
         Setup & — & w/o explan. & w/ explan. \\
         \midrule
         1M in. tokens & \$2 & \multirow{2}{*}{\$0.0065-\$0.0108} & \multirow{2}{*}{\$0.0234-\$0.0391} \\
         1M out. tokens & \$8 &  &  \\
         \bottomrule
    \end{tabular}
    \caption{{\color{black}Comparison of cost per 1 million input and 1 million output tokens for \texttt{GPT-4.1} and hosted \texttt{Gemma-3-12B-IT-RLV-FP8} on an \textbf{external} GPU provider in the label-only- and explanation-generation setups.}}
    \label{tab:cost_comparison}
\end{table}

\end{document}